\documentclass[numberedappendix]{emulateapj}

\usepackage{multirow}
\usepackage{color}

\begin{document}

\title{The Effect of Weak Lensing on Distance Estimates from Supernovae}
\author{
Mathew Smith\altaffilmark{*,1,2}, David J. Bacon\altaffilmark{3}, Robert C. Nichol\altaffilmark{3},  Heather Campbell\altaffilmark{3}, Chris Clarkson\altaffilmark{4}, Roy Maartens\altaffilmark{1,3}, Chris B. D'Andrea\altaffilmark{3}, Bruce A. Bassett\altaffilmark{2,5}, David Cinabro\altaffilmark{6}, David A. Finley\altaffilmark{7}, Joshua A. Frieman\altaffilmark{7,8,9}, Lluis Galbany\altaffilmark{10,11}, Peter M. Garnavich\altaffilmark{12}, Matthew D. Olmstead\altaffilmark{13}, Donald P. Schneider\altaffilmark{14,15} Charles Shapiro\altaffilmark{16} \& Jesper Sollerman\altaffilmark{17}}

\altaffiltext{*}{\email{matsmith2@gmail.com}}
\altaffiltext{1}{Department of Physics, University of the Western Cape, Cape Town, 7535, South Africa}
\altaffiltext{2}{South African Astronomical Observatory, P.O. Box 9, Observatory 7935, SA}
\altaffiltext{3}{Institute of Cosmology and Gravitation, University of Portsmouth, Portsmouth, PO1 3FX, UK}
\altaffiltext{4}{Astrophysics, Cosmology and Gravity Centre (ACGC), Department of Mathematics and Applied Mathematics, University of Cape Town, Rondebosch, 7701, SA}
\altaffiltext{5}{African Institute for Mathematical Sciences, 6-8 Melrose Road, Muizenberg 7945, SA}

\altaffiltext{6}{Wayne State University, Department of Physics and Astronomy, Detroit, MI 48202, USA}
\altaffiltext{7}{Center for Particle Astrophysics, Fermi National Accelerator Laboratory, P.O. Box 500, Batavia, IL 60510, USA}
\altaffiltext{8}{Kavli Institute for Cosmological Physics, The University of Chicago, 5640 South Ellis Avenue, Chicago, IL 60637, USA}
\altaffiltext{9}{Department of Astronomy and Astrophysics, The University of Chicago, 5640 South Ellis Avenue, Chicago, IL 60637, USA}
\altaffiltext{10}{CENTRA Centro Multidisciplinar de Astrof\'isica, Instituto Superior T\'ecnico, Av. Rovisco Pais 1, 1049-001 Lisbon, Portugal}
\altaffiltext{11}{Institut de F\'isica d'Altes Energies, Universitat Aut\`onoma de Barcelona, E-08193 Bellaterra (Barcelona), Spain}
\altaffiltext{12}{Department of Physics, University of Notre Dame, Notre Dame, IN 46556}
\altaffiltext{13}{Department of Physics and Astronomy, University of Utah, Salt Lake City, UT 84112}  

\altaffiltext{14}{Department of Astronomy and Astrophysics, The Pennsylvania State University, University Park, PA 16802}
\altaffiltext{15}{Institute for Gravitation and the Cosmos, The Pennsylvania State University,  University Park, PA 16802}
\altaffiltext{16}{NASA Jet Propulsion Laboratory; California Institute of Technology}
\altaffiltext{17}{The Oskar Klein Centre, Department of Astronomy, AlbaNova, SE-106 91 Stockholm, Sweden}

\shorttitle{Weak Lensing of SNe Ia}
\shortauthors{Smith et al.}

\begin{abstract}
Using a sample of 608 Type Ia supernovae from the SDSS-II and BOSS surveys, combined with a sample of foreground galaxies from SDSS-II, we estimate the weak lensing convergence for each supernova line-of-sight. We find that the correlation between this measurement and the Hubble residuals is consistent with the prediction from lensing (at a significance of $1.7\sigma$). Strong correlations are also found between the residuals and supernova nuisance parameters after a linear correction is applied. When these other correlations are taken into account, the lensing signal is detected at $1.4\sigma$. We show for the first time that distance estimates from supernovae can be improved when lensing is incorporated by including a new parameter in the SALT2 methodology for determining distance moduli. The recovered value of the new parameter is consistent with the lensing prediction. Using CMB data from WMAP7, $H_0$ data from HST and SDSS BAO measurements, we find the best-fit value of the new lensing parameter and show that the central values and uncertainties on $\Omega_m$ and $w$ are unaffected. The lensing of supernovae, while only seen at marginal significance in this low redshift sample, will be of vital importance for the next generation of surveys, such as DES and LSST, which will be systematics dominated.
\end{abstract}
\keywords{cosmology: observations --- distance scale -- supernovae: general --- surveys}

\section{Introduction}
\label{section_intro}

Type Ia supernovae (SNe Ia) are currently the best cosmological ``standard candles" and can be observed to high redshift. Extensive searches for SNe Ia have been carried out over the last decade to map the expansion history of the Universe with cosmic time. Observations of SNe Ia have produced convincing evidence that the Universe has undergone a recent period of accelerated expansion \citep{1998AJ....116.1009R,1999ApJ...517..565P,2006A&A...447...31A,2009ApJS..185...32K,2010MNRAS.401.2331L,2011ApJ...737..102S} leading to the inference that the energy density of the Universe is dominated by ``dark energy". By combining measurements of SNe Ia distances, over a wide range of redshift, with other cosmological probes such as measurements of the Cosmic Microwave Background (CMB) and Baryon Acoustic Oscillations (BAO), the equation of state of dark energy is known to an accuracy of $7\%$ \citep{2011ApJ...737..102S} and is consistent with a cosmological constant. 

While SNe Ia have been calibrated as ``standard candles", their luminosities retain a significant scatter about the best-fitting cosmological model, indicating that they are influenced by additional effects such as extinction (circumstellar and/or host galaxy dust), differences in the SN Ia progenitor, photometric calibration and possibly gravitational lensing. These systematic uncertainties can increase the dispersion of the SN Ia population's luminosity, and reduce the precision of the inferred constraints on cosmological parameters. Recent results from the Planck satellite \citep{2013arXiv1303.5076P} show some tension between the value of $\Omega_m$ determined by Planck and the most recent SN Ia datasets, suggesting that there could be residual systematic errors in either the SNe data, the Planck data or both that are not properly accounted for. Understanding and correcting for systematic uncertainties will be important in order to deliver the expected improvement in dark energy constraints from forthcoming surveys, such as the Dark Energy Survey \citep{dessn} and LSST \citep{2009arXiv0912.0201L}, which will observe thousands of SNe to high redshift. 

Recent progress in improving the standardization of SNe Ia has focussed on correlations between host galaxy properties and the observed SN parameters \citep{2010ApJ...715..743K,2010ApJ...722..566L,2010MNRAS.406..782S}. A strong correlation has been observed between the absolute magnitude of SNe Ia and the total stellar mass of the host galaxy. Applying this observed correlation does help reduce the scatter on the Hubble diagram, thus improving the cosmological parameter estimates \citep{2011ApJ...737..102S}. However, the origin of this empirical correction remains unclear \citep{2011ApJ...743..172D,2011ApJ...740...92G,2012ApJ...755..125G,2013ApJ...764..191H} and may evolve with redshift. 

One expected and well-understood cause for an increase in the dispersion of SN Ia magnitudes is the weak gravitational lensing of SN light by the intervening matter along the line of sight \citep{1996ComAp..18..323F,1998ApJ...506L...1H,1998ApJ...507..483K,1999MNRAS.305..746M,1999ApJ...525..651W,2000A&A...358...13B,2003A&A...397..819A,2003A&A...403..817M,2005ApJ...631..678H,2006PhRvL..96b1301C,2010PhRvL.105l1302A,2012MNRAS.426.1121C,2013arXiv1304.7689M,2013PhRvD..87l3526F}. Correlations between the background SNe (point sources) and the foreground clustered mass will cause SNe Ia to appear brighter, relative to the mean of the SN Ia population, when the lensing convergence along the line of sight is positive, and conversely de-magnified when it is negative. (Note that there is an additional Doppler contribution to the convergence which acts in the opposite sense and can be significant at low redshifts \citep{2013PhRvL.110b1302B}.) This effect will not significantly bias the cosmological parameters \citep{2008ApJ...678....1S,2008A&A...487..467J}, but if uncorrected, will cause additional scatter in the observed SN Ia magnitudes leading to an increase in their distance uncertainty. This additional dispersion is greater at high redshifts due to the additional extent of the light-path. \citet{2000ApJ...531..676W} used simulated data with weak lensing noise to show that the estimated cosmological parameters will be unbiased if the fitting is carried out in flux space, suggesting a flux-averaging approach as lensing conserves total flux. 

Several studies have addressed the expected increased dispersion in SN Ia magnitudes due to weak gravitational lensing. \citet{1997ApJ...475L..81W} considered ray--tracing in cosmological simulations and found an increased scatter of $0.02$ mag for SNe Ia at $z=0.5$, while \citet{2003A&A...403..817M} and \citet{2003MNRAS.346..949T} showed that lensing causes variations of $\delta m \lesssim 0.01$.  These effects are presently sub-dominant to the current SN Ia magnitude uncertainties of $\delta m\simeq 0.15$ mag. \citet{1996ComAp..18..323F} shows that for sources at $z=1$, density fluctuations could increase the observed dispersion by $30\%$. \citet{2006ApJ...640..417G} showed that for an SN Ia at $z=1.5$, the dispersion due to lensing is comparable to the intrinsic SN Ia scatter, and introduced a 
method to reduce the scatter from $7\%$ to $3\%$.

The effect of lensing on high redshift SN Ia datasets has been studied by several authors. \citet{2004MNRAS.351.1387W} used a sample of 55 SNe Ia with $z\ge 0.35$ from the Supernova Cosmology Project and High-z Supernova Search datasets \citep{2003ApJ...594....1T} and correlated their brightness with foreground galaxies from the APM Northern Sky Catalogue \citep{1994IEEES...2...14I}. They detected a correlation consistent with lensing at the $>99\%$ confidence level, but the observed difference of $0.3$ mag, between the most magnified and de--magnified SNe Ia, is far larger than expected.  \citet{2005MNRAS.358..101M} used 44 SNe Ia from the \citet{2004ApJ...607..665R} dataset in combination with galaxies from the photometric catalogue of the Sloan Digital Sky Survey (SDSS) and found no detectable correlation on scales of one to ten arcminutes. \citet{2005JCAP...03..005W} convolved the intrinsic distribution of SNe Ia, using the \citet{2004ApJ...607..665R} sample, with magnification distributions of point sources, finding marginal evidence for a non-Gaussian tail at high redshift, and a shift in the peak brightness towards the faint end, both indicators of weak lensing. \citet{2001ApJ...561..106M} considered SN 1997ff at $z=1.77$ and showed that careful modeling of foreground galaxies is required to estimate the lensing signal, finding a large range of possible magnifications. \citet{Jonsson:2006eu} found a signal consistent with lensing at $\sim90\%$ confidence level using 26 SNe in the GOODS field and an aperture of one arcminute to estimate the foreground galaxy density.

Recently, several authors have looked for lensing using the larger, more homogeneous, three-year data release of the Supernova Legacy Survey (SNLS) \citep{2006A&A...447...31A}. \citet{2010A&A...514A..44K} combined this dataset with a deep photometric catalogue of foreground galaxies (with inferred masses) to find evidence for a lensing signal at $2.3\sigma$, while \citet{2010MNRAS.405..535J} detected a signal at $92\%$ confidence, simultaneously constraining the properties of the galaxy dark matter haloes. However, \citet{2012arXiv1207.3708K} obtained only a marginal detection of a lensing signal when using a Bayesian analysis of the same dataset, and only found weak constraints on the dark matter halo parameters. 

The expected lensing contribution to the scatter of SN Ia magnitudes is not anticipated to be strong for current SN Ia samples, due to the small number of confirmed SNe Ia, the limited redshift range surveyed and photometric uncertainties. However, with future surveys, such as DES and LSST producing thousands of SNe Ia to $z>1$, the gravitational lensing effect should become important, especially to achieve the required high precision on the cosmological parameters. It is therefore important to develop a model-independent formalism to characterize and account for this effect. 

In this paper, we develop a scheme to measure and correct for the effect of weak gravitational lensing on Type Ia supernova distances. Using a  new sample of 608 SNe Ia obtained from the SDSS-II SNe Survey, with $0.2 < z < 0.6$, supplemented by spectroscopic host galaxy redshifts observed as part of the BOSS survey \citep{2011AJ....142...72E,2012arXiv1208.0022D}, we correlate this SN sample with foreground galaxies from SDSS-II to constrain the possible lensing signal. We also extend previous analyses by simultaneously constraining the lensing signal alongside other SN nuisance parameters, thus improving the standardization of SNe Ia. 

The outline of this paper is as follows. In \S\ref{sec:data} we describe the SN Ia and galaxy data used in this analysis. \S\ref{sec:estimatingkappa} describes the analysis used to estimate the lensing signal, while in \S\ref{sec:results}, we present the measured correlation, and its impact on the cosmological parameters. In \S\ref{sec:snedistances}, we discuss how the lensing signal will affect the inferred distances to SNe Ia and constrain the bias of our foreground galaxy sample. Finally, we conclude in \S\ref{sec:conclusions}.

\section{Data}
\label{sec:data}

\subsection{The SDSS-II Supernova Survey}
\label{subsec:snedata}

From 2005 to 2007, the SDSS-II SN Survey \citep{2008AJ....135..338F,2008AJ....135..348S} carried out a dedicated search for intermediate-redshift SNe Ia from repeated scans of the equatorial ``Stripe82" region covering a total of $300\, \textrm{deg}^2$. The SDSS 2.5m Telescope \citep{1998AJ....116.3040G} carried out multi-colour \textit{ugriz} imaging for three months a year (September to November) with an average cadence of 3 days. Using a suite of international telescopes \citep{2008AJ....135.1766Z,2011A&A...526A..28O,2011arXiv1101.1565K}, over 500 SNe Ia were spectroscopically confirmed, with several thousand additional probable SNe Ia identified through their high quality light-curves. 

This sample of SDSS-II SNe Ia has now been used to constrain cosmological parameters \citep{2009ApJS..185...32K,2010MNRAS.401.2331L,2009ApJ...703.1374S}, measure the SN Ia rate \citep{2008ApJ...682..262D,2010ApJ...713.1026D,2012ApJ...755...61S}, examine the rise-time distribution \citep{2010ApJ...712..350H} and study the correlation between SNe Ia and their host galaxies \citep{2010ApJ...722..566L,2011ApJ...743..172D,2011ApJ...740...92G, 2012ApJ...755..125G,2013ApJ...764..191H} and spectroscopic indicators \citep{2011ApJ...734...42N,2011A&A...526A.119N,2011arXiv1103.2497K,2012AJ....143..113F}. 

\subsection{BOSS Ancillary Program}

In 2009, an ancillary program was initiated as part of the SDSS-III Baryon Oscillation Spectroscopic Survey (BOSS) \citep{Olmstead2013,2012arXiv1208.0022D} to obtain the spectra and redshifts of the host galaxies of a large sample of supernova candidates detected by the SDSS-II SN Survey. This program was designed to understand possible incompletenesses and biases in the original real-time spectroscopic follow-up. In total, spectra were obtained for 3761 host galaxies, producing 3520 confirmed redshifts of SN candidates (and other transients) to a limiting galaxy magnitude of $r<22.0$. Full details of the target selection and data reduction for this sample of galaxies can be found in \citet{2013ApJ...763...88C}, while details of the data analysis and redshifts for the sample are presented in \citet{Olmstead2013}.

The details of how a robust cosmological Hubble diagram is constructed using these host galaxy redshifts in conjunction with the original SDSS-II SN light curve data is presented in \citet{2013ApJ...763...88C}. Using a combination of the PSNID \citep{2011ApJ...738..162S} and SALT2 \citep{2007A&A...466...11G} techniques, combined with stringent data-quality cuts,  a new and robust sample of 752 photometrically classified SNe Ia covering a redshift range $0.05<z<0.55$ was constructed from the BOSS and SDSS-II galaxy samples. Using realistic simulations, \citet{2013ApJ...763...88C} showed that this sample is over 70\% efficient in detecting SNe Ia over this redshift range, with only 4\% probable contamination from non-Ia supernovae. \citet{2013ApJ...763...88C} further demonstrated that this sample provides competitive cosmological constraints, compared to the spectroscopically confirmed samples from SNLS. 

For this work, the distance modulus to an SN Ia is determined using the SALT2 light-curve fitting method \citep{2007A&A...466...11G} and is defined as 

\begin{equation}
\mu = m_B - M + \alpha x_1 - \beta c + \mu_{\mathrm{corr}} (z),
\label{eq:basic_salt}
\end{equation}

\noindent where $m_B = 10.635 - 2.5\log x_0$. $x_0$, $x_1$ and $c$ are SNe parameters determined through fitting of the individual light-curves, and correspond to the peak magnitude, stretch and color of each SN. Here $M$ is the absolute peak magnitude of a standard SNe Ia (assumed to be $-19.0$ for this analysis) and $\alpha$ and $\beta$ are global SALT2 parameters that describe the relationship between the stretch and colour of an SN Ia and the absolute brightness. We also include a correction for Malmquist bias ($\mu_{\mathrm{corr}}(z)$) which is discussed, and calculated for this sample, in \citet{2013ApJ...763...88C}. 

For our fiducial analysis, we use $\alpha=0.22, \beta=3.12$ and the best-fitting cosmology taken from \citet{2013ApJ...763...88C} of $(\Omega_{m}, \Omega_{\Lambda}, \Omega_{k}, w) = (0.27, 0.73, 0.0, -0.95)$ and $H_0 = 73.8\textrm{km}\,\textrm{s}^{-1}\,\textrm{Mpc}^{-1}$ from the SH0ES survey \citep{SHOES}. As with other SN analyses, we also include an intrinsic dispersion for the sample of $\sigma_\mathrm{int}=0.12$ mag, which provides a reduced $\chi^2$ close to unity for the best fit.  To remove possible bias from large outliers, which can significantly impact any correlation, a $5\sigma$ clip on residuals from the Hubble diagram using the best-fitting cosmological parameters, has been applied to the data, removing three SNe Ia in total, reducing our sample to 749 SNe Ia. 

\begin{figure*}[t]
\epsscale{1.0}
\plotone{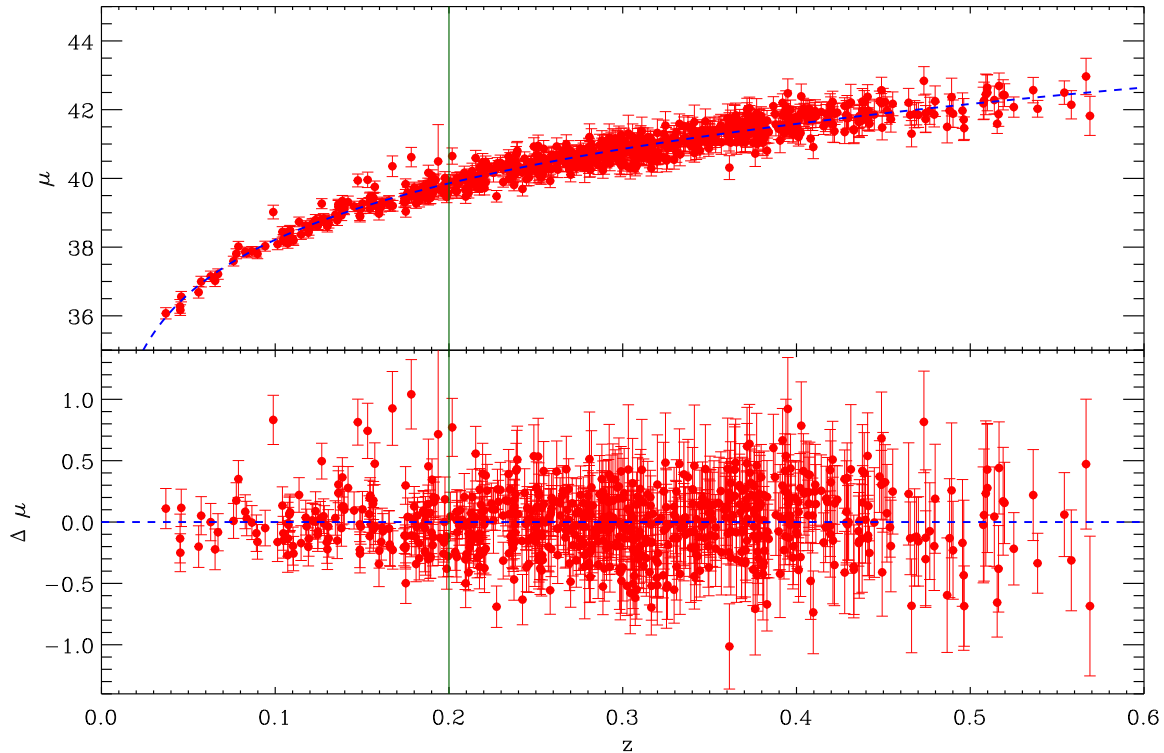}
\caption{\emph{Upper:} Hubble diagram for the 749 SNe used in this analysis. \emph{Lower:} Residuals from the above Hubble diagram, considering the best-fitting cosmology from \citet{2013ApJ...763...88C}. The redshift cut of $z>0.2$, used to define the fiducial sample, is also shown.}
\label{fig:hubble}
\end{figure*}

Figure~\ref{fig:hubble} gives the Hubble diagram for the 749 photometrically classified SNe Ia taken from \citet{2013ApJ...763...88C} and Figure~\ref{fig:redshift_hist} shows the redshift histogram for these SNe Ia. Our fiducial sample consists of 608 SNe Ia with $0.2 < z < 0.6$. This selection is discussed in \S\ref{sec:estimatingkappa}. This is one of the largest SNe Ia datasets in existence and is appropriate for the lensing study discussed in this paper because of the uniform selection, consistent relative photometric calibration (all of the SNe Ia data is from a single survey) and high completeness. The sample also pushes to higher redshift than the spectroscopically confirmed SNe sample of SDSS-II, which helps in our search for a gravitational lensing signal.  

\subsection{Spectroscopic Galaxy Catalogue} 
\label{subsec:galdata}

In addition to background SNe, we need tracers of the foreground mass density in order to obtain a lensing correlation between SN brightness and foreground density. Ideally, these foreground tracers would have accurate spectroscopic redshifts allowing us to unambiguously determine their location relative to the SNe. This also facilitates a better prediction of the expected lensing signal taking into account the relative distances between us, the SNe and foreground lenses. Fortunately the ``Stripe82" region of the SDSS has a significantly higher density of spectroscopic data compared to the average for SDSS due to a number of other ancillary programs as outlined in the Data Release 8 (DR8) of the SDSS \citep{2011ApJS..193...29A}. In total, there are over 800,000 spectroscopic galaxy redshifts in DR8 in the ``Stripe82" region.  

\begin{figure*}[t]
\epsscale{1.0}
\plotone{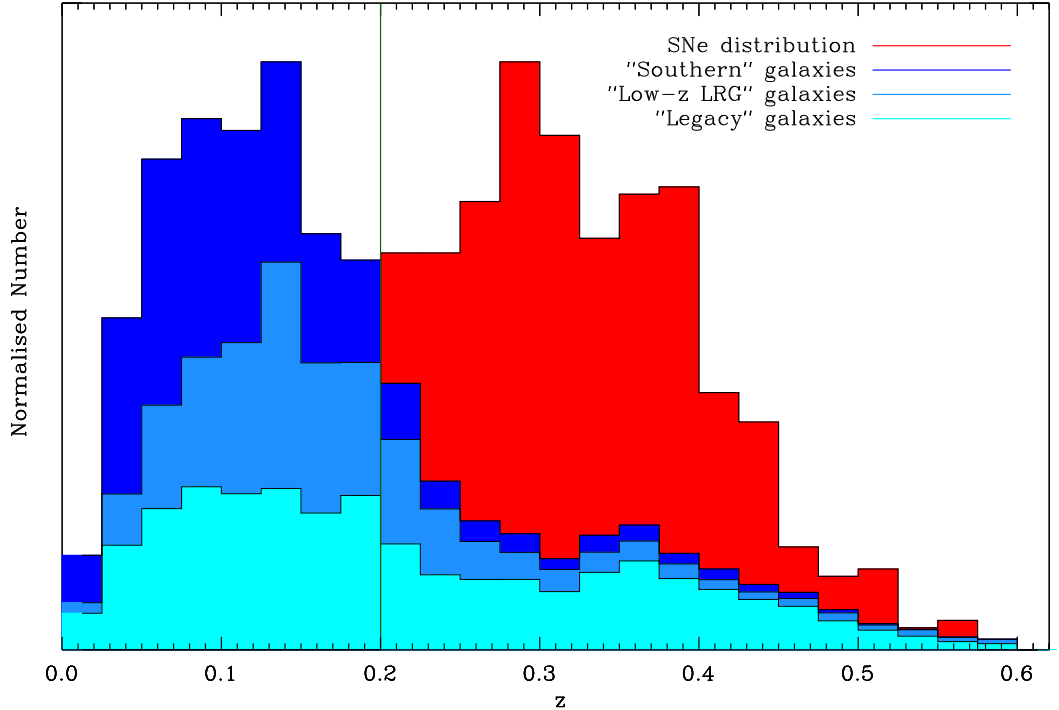}
\caption{Normalized redshift distribution for the 749 SNe Ia (red) and 70631 galaxies used in this analysis, as described in \S\ref{sec:data}. The redshift histograms for the three galaxy sub-samples are shown cumulatively. }
\label{fig:redshift_hist}
\end{figure*}

We have not used all these galaxy redshifts, but instead use only the ancillary programs with well-defined selection criteria that span the whole area of ``Stripe82". In this way, we can be more confident of the homogeneity of the selection, which is important for studying the expected small lensing signal. First, we use galaxies selected by the standard {\it SDSS-I/II Legacy Survey}, namely the Main Galaxy Sample (MGS; \citealt{2002AJ....124.1810S}) consisting of 22918 galaxies. In addition, we use a sample of 19589 galaxies from the {\it Low-z LRG} program, which carried out a survey of low-redshift galaxies to two magnitudes fainter than the MGS in order to add more low-luminosity galaxies to the MGS, and included a deeper sample of LRGs and Brightest Cluster Galaxies. Finally, we use 28124 galaxies from the {\it Southern} program, executed on the Equatorial stripe in the Southern Galactic Cap, designed to create a region of the sky where the MGS is close to $100\%$ complete. Details of these ancillary programs can be found in \citet{2011ApJS..193...29A} and on the DR8 webpage$\footnote[1]{\tt http://www.sdss3.org/dr8/algorithms/special\_target.php}$. Together, these sub-samples combine to give 70,631 galaxies with a spectroscopic redshift spread over the``Stripe82" region. We show the redshift histogram of these different samples in Figure~\ref{fig:redshift_hist}. 

\section{Estimating the Lensing Signal}
\label{sec:estimatingkappa}

In this section we describe the estimator used to predict the expected lensing signal for a given SN, at redshift $z$, and discuss the  robustness of this estimator. 

Assuming a flat Universe ($\Omega_k = 0$), the \textit{convergence}, $\kappa$ for a source on a particular line-of-sight can be approximated by,

\begin{equation}
\kappa = \frac{3 {H_0}^2 \Omega_{m}}{2 c^2} \sum_i \Delta_{\chi_i}\, {\chi_i} \frac{(\chi_\mathrm{SN} - {\chi_i})}{\chi_\mathrm{SN}} \frac{\delta_i}{a_i},
\label{eq:lensing}
\end{equation}

\noindent where the matter distribution along the line-of-sight is binned into shells in redshift ($z_i$) with corresponding co-moving distances of $\chi_i$  and bin width $\Delta_{\chi_i}$ \citep{2001PhR...340..291B}. Here $H_0$ is the Hubble constant, $\Omega_m$ is the matter density parameter, $c$ is the speed of light, $a_i$ is the scale factor for bin $i$, and $\delta_i$ is the overdensity of matter in the $i$th bin. The source co-moving distance is given by ${\chi_\mathrm{SN}}$; in our case, this source is a SN Ia. A line-of-sight with $\kappa>0$ should, on average, result in a brightening of a SN Ia.

Equation~\ref{eq:lensing} predicts the convergence based on the true matter distribution along a given line-of-sight. In \S\ref{subsec:galdata}, we introduced a sample of galaxies that can be used to trace that matter distribution. This distribution can be considered a sample of point sources that, when smoothed, approximates the underlying matter distribution. We estimate $\kappa$ by replacing $\delta_i$ in Equation \ref{eq:lensing} with $\delta_{n_i} = [{n(z_i) - \bar{n}(z_i)}]/{\bar{n}(z_i)}$, which is the galaxy over-density in a given redshift bin compared to the mean number density in that redshift bin ($\bar{n}(z_i)$), determined from our sample of 70,631 galaxies spanning the ``Stripe82" region.  However, any sample of galaxies will be biased with respect to the distribution of dark matter. Assuming a linear bias $b$ between the galaxy sample and the underlying distribution of matter, we can relate the true value of $\kappa$ to that measured through the galaxy distribution, using $\kappa= \kappa_\mathrm{gal}/b$.

The uncertainty on $\kappa_\mathrm{gal}$ due to Poisson noise is

\begin{equation}
\sigma_{\kappa}^2 = \frac{3 {H_0}^2 \Omega_{m}}{2 c^2}  \sum_i \Delta_{\chi_{i}} \, \chi_i \frac{(\chi_\mathrm{SN} - {\chi_i})}{\chi_\mathrm{SN}} \frac{1}{a_i}\times \frac{1}{\bar{n}_i}.
\label{eq:lensing_err}
\end{equation}

\begin{figure*}[t]
\epsscale{1.0}
\plotone{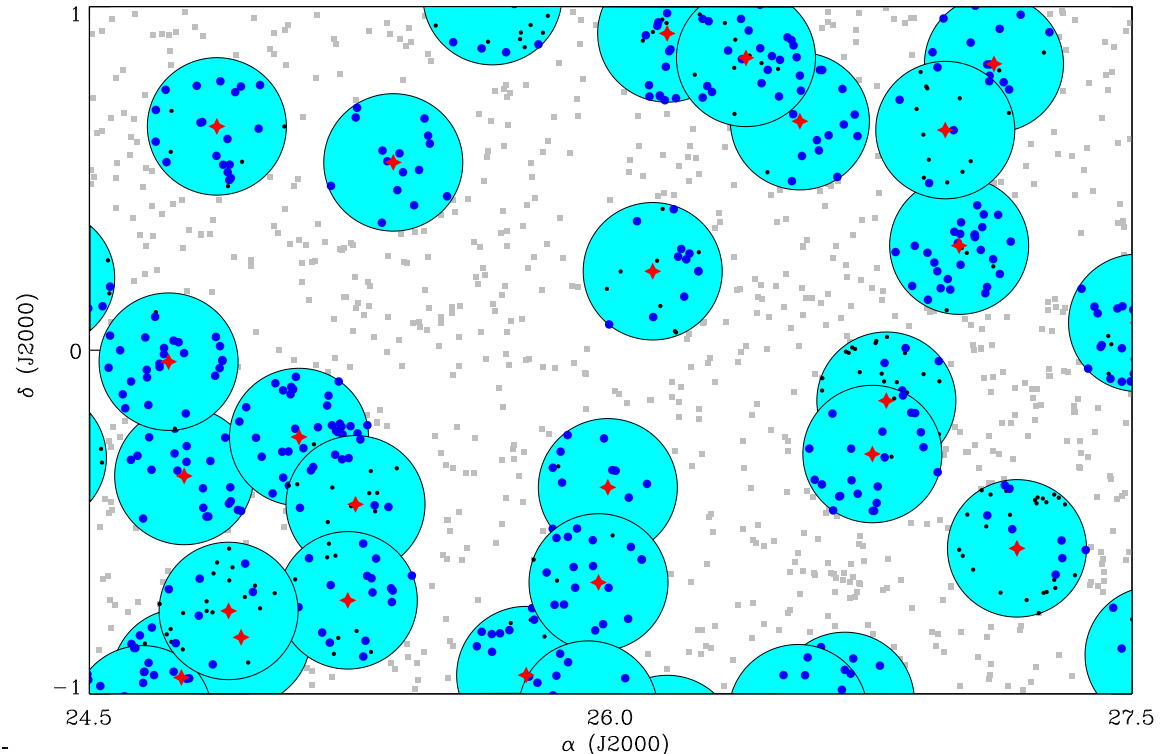}	 
\caption{An illustration of the methodology used to search for SN lensing in our SDSS SN sample. We only show a small portion of the ``Stripe82" field and highlight with stars the SDSS SNe. Around each SN, we show the projected 12 arcminute aperture used to calculate $\kappa_\mathrm{gal}$ in cyan. We show galaxies within an aperture and in the foreground of a SN as blue dots. Galaxies within an aperture, but behind the SNe are shown in black. 
\label{fig:example}} 
\end{figure*}

In practice, we must determine $\kappa_\mathrm{gal}$, for a given SN light--of--sight, using an aperture centered on each background SN. This methodology is illustrated in Figure~\ref{fig:example} where we show a portion of the ``Stripe82" area and highlight the SNe and galaxies in this region. We use a 12 arcminute radius aperture around each SN which we have determined to be the optimal radius for our measurement. In Appendix~\ref{app:angular}, we show that the choice between angular or cylindrical apertures does not substantially affect our measurement, while in Appendix~\ref{app:opt_ap} we show how our results depend on the choice of aperture considered. We re-scale the number density of galaxies in apertures that fall partially outside the boundaries of ``Stripe82" . Unless otherwise stated, we assume angular apertures of 12 arcminutes throughout this paper. \citet{2003ApJ...585L..11D} find that a considerable fraction of the lensing dispersion derives from subarcminute scales caused by the substantial small-scale power in the mass distribution at these scales. Our estimate of $\kappa_\mathrm{gal}$, while not probing these scales, is a smoothed estimate of the underlying $\kappa$ distribution. 

\begin{figure*}[t]
\epsscale{1.0}
\plotone{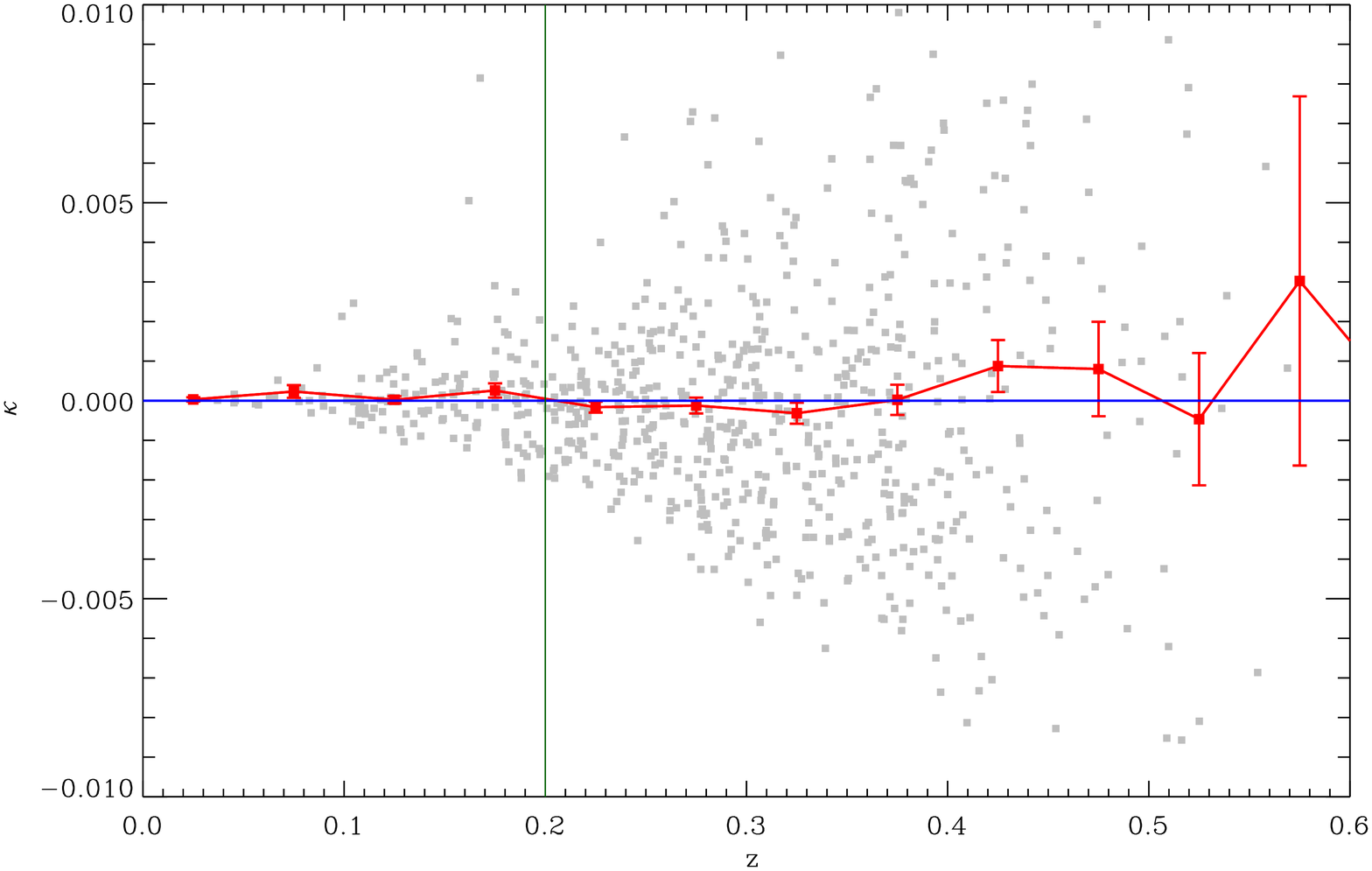}
\caption{The distribution of $\kappa_\mathrm{gal}$ as a function of redshift for the SDSS SN sample, when a fixed aperture of $12$ arcminutes is considered.  The mean values of $\kappa$ in bins of $\delta z =0.025$ are shown in red. The redshift cut of $z>0.2$, used to define the fiducial sample, is given by the green vertical line.}
\label{fig:kappa_v_redshift}
\end{figure*}

\begin{figure*}[t]
\epsscale{1.0}
\plotone{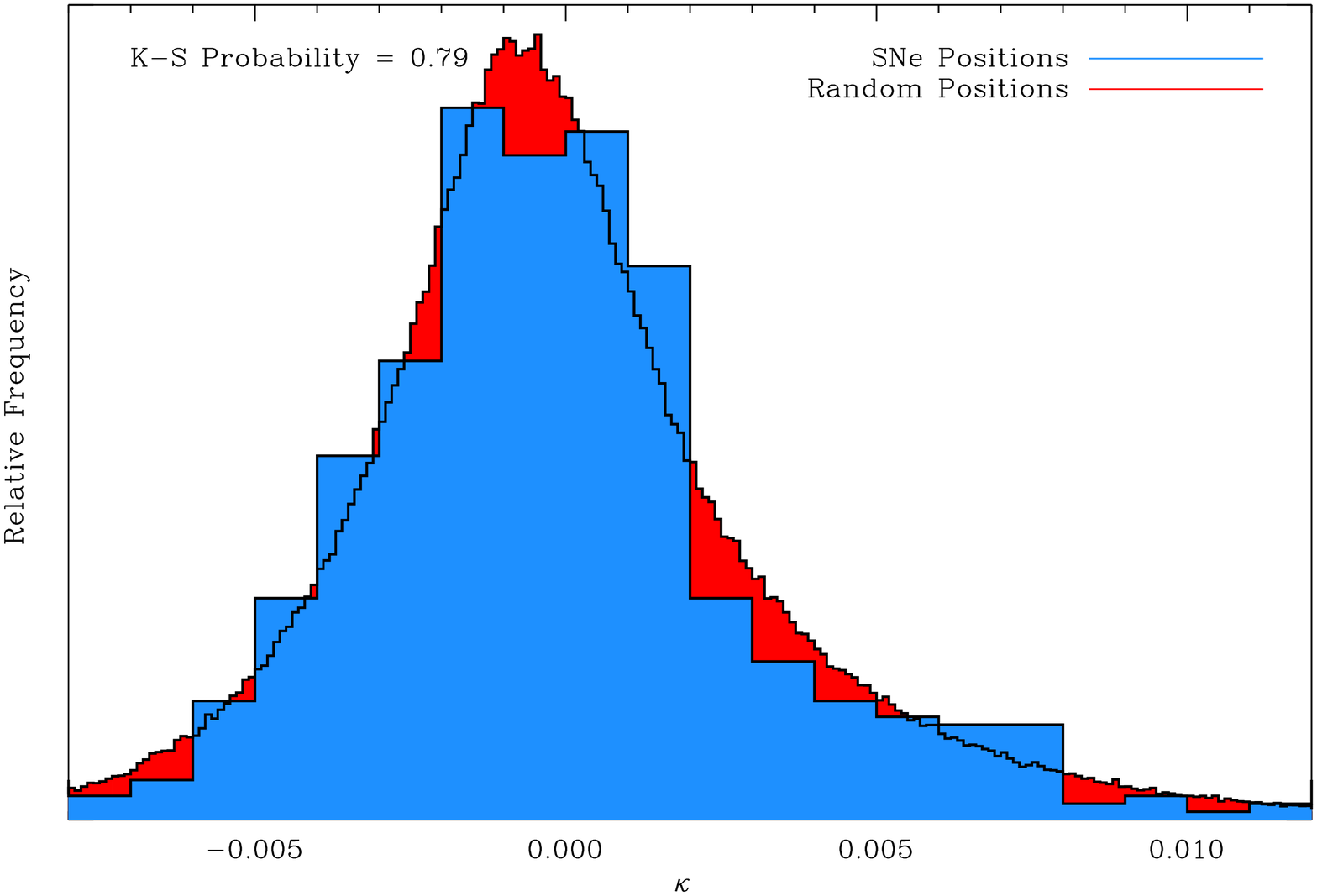}	
\caption{Normalized histogram of the distribution of $\kappa_\mathrm{gal}$ for the sample of 749 SDSS SNe (in blue), compared to a sample produced from 1000 realizations of 749 random positions within the ``Stripe82" footprint (shown in red), when a fixed aperture of $12$ arcminutes is considered.  The probability obtained from a K-S test is also shown.
\label{fig:kappadist}}
\end{figure*}

For random lines of sight, \citet{1976ApJ...208L...1W} shows that at a fixed redshift, the mean convergence $\bar{\kappa}$ is zero and the dispersion $\sigma_\kappa$ increases with increasing redshift. To test whether our SN Ia positions are consistent with random lines-of-sight, we show in Figure~\ref{fig:kappa_v_redshift} the distribution of $\kappa_\mathrm{gal}$ as a function of redshift for the 749 SNe Ia in our sample and the SDSS foreground galaxies. We observe that, on average, at a fixed redshift, $\bar{\kappa}=0$, while $\sigma_\kappa$ increases with redshift. To further test this hypothesis, we randomized the SN positions within the ``Stripe82" region, and repeated the measurement. Figure~\ref{fig:kappadist} shows a normalized histogram of the distribution of $\kappa_\mathrm{gal}$ for the 749 SNe Ia compared to 1000 realizations of 749 random positions within the ``Stripe82" region. The randomized positions have $\sigma_\kappa=3.65\times{10}^{-3}$ mag, consistent with the 749 SN Ia positions, which have $\sigma_\kappa=3.56\times{10}^{-3}$ mag. A Kolmogorov-Smirnov test (K-S test, see \citealt{chakravartistat}) comparing these distributions gives a probability of $0.79$, indicating that our SN Ia positions are likely not different from those of random lines of sight.

Figure~\ref{fig:kappa_v_redshift} shows that the dispersion on the convergence increases with increasing redshift. Since SNe with $z < 0.2$ do not have a significant $\kappa_\mathrm{gal}$ along their lines of sight, due to the limited volume probed at these redshifts, we only consider supernovae with $z>0.2$ for our analysis, reducing our sample to 608 SNe Ia.  The implications of this cut are discussed in Appendix~\ref{app:zmin}.

\section{Results}
\label{sec:results}

\subsection{Correlating galaxy density with SN Hubble residuals}
\label{subsec:fiducialresult}

\begin{figure*}[t]
\epsscale{1.0}
\plotone{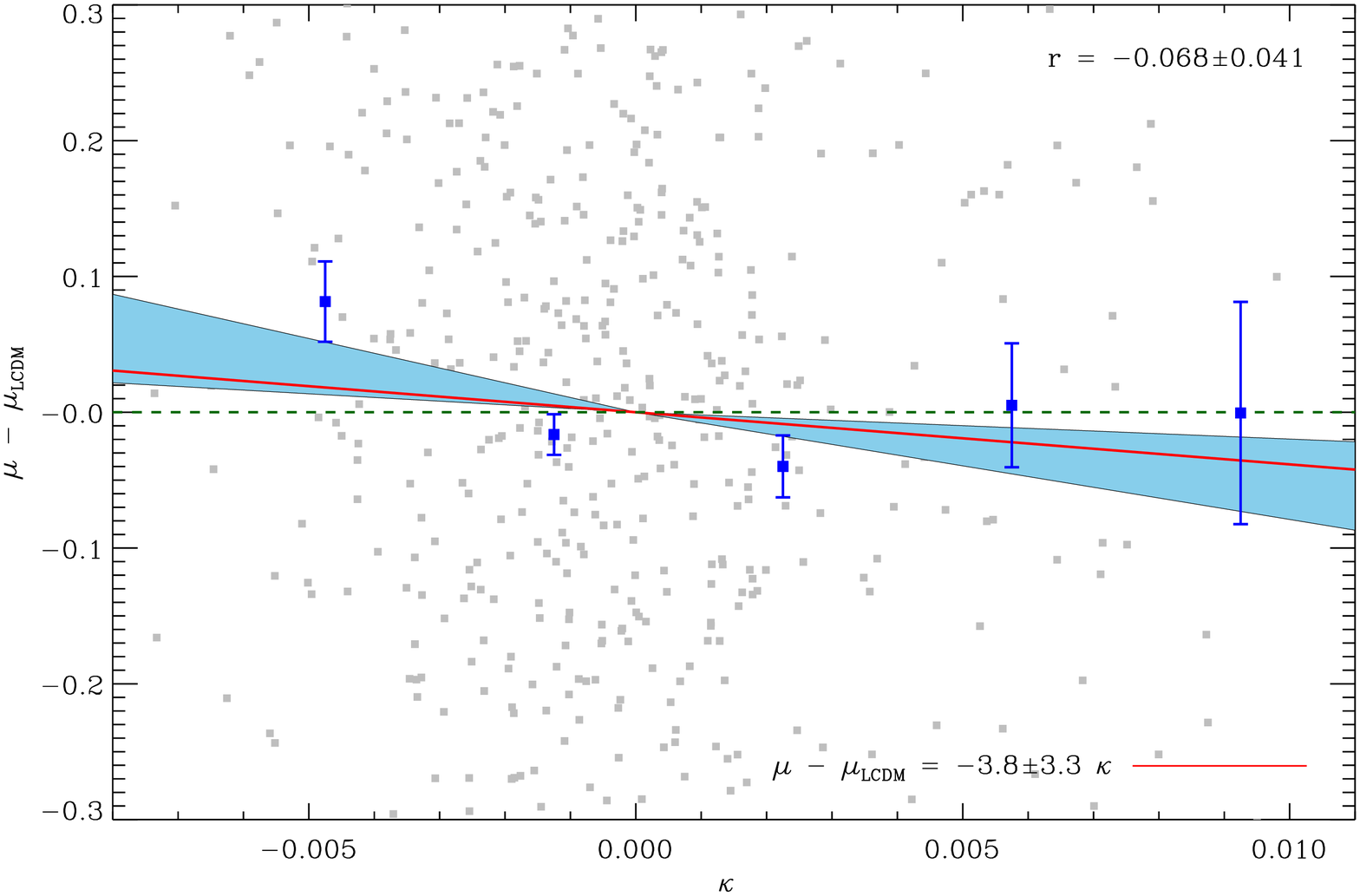}		
\caption{$\kappa_\mathrm{gal}$ when a fixed angular radius of $12$ arcminutes is considered compared to residuals from the Hubble diagram for the 608 SNe Ia in our sample, shown in grey. A line of best fit is shown in red, while the correlation between the two quantities is also given. The anticipated correlation for our sample, assuming a conservative range of $b=0.5$ to $2$ is shown in light blue. The mean values in bins of $\kappa_\mathrm{gal}$ are shown in light blue. The case of no correlation is shown as a green dashed line. 
\label{fig:kappa_v_resid}}
\end{figure*}

Figure~\ref{fig:kappa_v_resid} shows, in grey, the observed correlation between the Hubble residuals ($\Delta\mu=\mu_\mathrm{obs} - \mu_\mathrm{cosmo}$) of our 608 SDSS SNe Ia and our estimate of $\kappa_{gal}$ along each line--of--sight (with a fixed angular aperture of $12$ arcminutes), assuming the fiducial cosmology described in \S\ref{sec:snedistances}. A dashed green line shown in Figure~\ref{fig:kappa_v_resid} indicates a null correlation between these two quantities, while the red line shows the best linear fit to the data. The mean values in bins of $\kappa_\mathrm{gal}$ are shown in blue, while the expected lensing signal, discussed in \S\ref{sec:snedistances}, assuming a conservative $b=[0.5,2]$ is shown as a blue band. 

We use the Spearman's rank correlation coefficient ($\rho$) to statistically determine the significance of any correlation seen in Figure~\ref{fig:kappa_v_resid}, i.e., the best-fit red line. For an aperture of 12 arcminutes, we find $\rho = -0.068 \pm 0.041$, which is a detection of a correlation of $1.7\sigma$. This significance is comparable to the $2.3\sigma$ correlation found by \citet{2010A&A...514A..44K} using 233 SNe Ia from the SNLS dataset and \citet{Jonsson:2006eu} using 26 SNe Ia from the GOODS field who found a tentative detection of lensing at $90\%$ confidence level. 
This limited significance correlation is consistent with the expected weak lensing signal for our data. SNe Ia with $\kappa_\mathrm{gal}>0$ have $\bar{\Delta\mu} = -0.024\pm0.017$ mag, compared to $\bar{\Delta\mu} = 0.010\pm0.014$ mag for those with $\kappa_\mathrm{gal}<0$. 

\subsubsection{Comparing to the expected lensing signal} 
\label{sec:lensingprediction}

To determine if the scatter we observe on $\kappa_\mathrm{gal}$ in Figures~\ref{fig:kappa_v_redshift} and ~\ref{fig:kappa_v_resid} is consistent with expectation, we estimate the theoretical error on $\kappa$, by considering $\kappa$ averaged in an angular aperture of radius $\theta$. Following the analysis of \citet{2001PhR...340..291B}, we calculate the weight function, $\bar{W}(\chi)$, defined as, 

\begin{equation}
\bar{W}(r)= \int^{\chi_H}_{\chi} d\chi'G(\chi')\frac{f_\kappa(\chi'-\chi)}{f_\kappa(\chi')},
\label{lensing_weighting}
\end{equation}

\noindent where $\chi$ is the comoving distance, and $G(\chi)d\chi=p_z(z)dz$. Assuming a flat Universe such that $f_k(\chi)=\chi$, we calculate $P_\kappa$, by integrating $\bar{W}$ and the power spectrum ($P_{\delta}$),

\begin{equation}
P_{\kappa}(l)=\frac{9 H_0^4 \Omega_0^2}{4c^2}\int_{0}^{w_H}d\chi \frac{\bar{W}^2(\chi)}{a^2(\chi)}P_{\delta}\left(\frac{l}{f_K(\chi)},\chi\right),
\label{lensing_P_K}
\end{equation} 

\noindent where $a$ is the scale factor. The $rms$ scatter on $\kappa$, within a circular aperture of radius $\theta$ is,

\begin{equation}
\langle \kappa^2_{av}(\theta)\rangle=2\pi \int^{\infty}_{0} l dl P_\kappa(l)\left[\frac{J_1(l\theta)}{\pi l\theta}\right]^2 ,
\label{lensing_S_N}
\end{equation} 

\noindent where $J_1(x)$ is the first order Bessel function of the first kind. Considering an aperture of 12', with $(\Omega_{m}, \Omega_{\Lambda}, w) = (0.3, 0.7, -1.0)$, $\sigma_8=0.8$ and $H_0 = 70\,\textrm{km}\,\textrm{s}^{-1}\,\textrm{Mpc}^{-1}$ we use the package {\it iCosmo} \citep{2011A&A...528A..33R} to calculate the non-linear power spectrum using the fitting formula of \citet{1994MNRAS.267.1020P} and determine $\langle \kappa^2_{av}(\theta)\rangle$ from Equations~\ref{lensing_weighting},~\ref{lensing_P_K} and~\ref{lensing_S_N}. We find an expected $rms$ scatter on $\kappa$ of 0.44$\%$ over the redshift range for this supernova sample, consistent with the value of $0.36\%$ for our dataset as seen in Figures~\ref{fig:kappa_v_redshift},~\ref{fig:kappadist} and~\ref{fig:kappa_v_resid}. 

\subsubsection{Determining the source and significance of the correlation}
\label{subsubsec:ppcctables}

The mild correlation seen in Figure~\ref{fig:kappa_v_resid} could be due to a systematic uncertainty or a manifestation of a different astrophysical effect rather than weak gravitational lensing, even though the sign of the observed effect is as expected from lensing (i.e. brighter residuals are seen along lines of sight with positive $\kappa_{gal}$). In particular, recent studies have shown that passive, more massive, galaxies host brighter SNe Ia even after light-curve correlation \citep{2010ApJ...722..566L}, and this could be responsible in part for the correlation seen in Figure~\ref{fig:kappa_v_resid} as such massive galaxies reside in high density regions (clusters) which themselves are highly clustered. 

To further test the correlation of $\Delta\mu$ and $\kappa_{gal}$, we also consider the supernova observables $x_1$ (light curve stretch) and $c$ (color) as part of our Spearman correlation coefficient analysis. In Table~\ref{tab:pearson2} we provide the Spearman's rank correlation coefficient, $\rho$, and the significance of any detection, for a suite of possible correlations between these SN observables and $\kappa_{gal}$. First, we see a strong correlation between $\Delta \mu$, $x_1$ and $c$, stronger than that seen between $\Delta \mu$ and $\kappa_\mathrm{gal}$. This correlation is observed after the supernova distances have been linearly corrected using the global parameters $\alpha$ and $\beta$, as in Equation~\ref{eq:basic_salt}, suggesting the possibility of non-linear corrections being required.  A  correlation at $2.2\sigma$ is observed between $\kappa_\mathrm{gal}$ and $x_1$, such that SNe on overdense lines of sight ($\kappa>0$) have the smaller values of $x_1$. 

\begin{deluxetable*}{cccc}
  \tablewidth{0pt}
  \tablecaption{Spearman coefficient, $\rho$ between various SN observables. The significance of each correlation is also given in brackets. \label{tab:pearson2}}
  \tablehead{
  \colhead{} &
  \colhead{c} &
  \colhead{$x_1$} &
  \colhead{$\kappa_\mathrm{gal}$}
  }
  \startdata
$\Delta \mu$ &  $-0.33 \pm 0.04 (8.1)$ &  $0.49 \pm 0.04 (12.0)$ &  $-0.07 \pm 0.04 (1.7)$ \\
c &  - & $0.10 \pm 0.04 (2.6)$ &  $-0.05 \pm 0.04 (1.1)$ \\
$x_1$ &  - & - &  $-0.09 \pm 0.04 (2.2)$ \\
\end{deluxetable*}

However, while this result indicates that there are significant correlations between the parameters that we are considering, it does not highlight the underlying source of them. To study this, we use the the \emph{Partial Correlation Coefficient}, $r$. This statistic determines the correlation between two variables when the effects of all other variables considered are removed. This statistic allows us to determine the residual correlation after the correlations between other parameters have been considered. Table~\ref{tab:pearson3} shows the value of $r$ (and its significance) between each of the SN observables considered in this analysis, when correlations between all of the other parameters have been removed. For example, from Table~\ref{tab:pearson2} a correlation between $\Delta \mu$ and $x_1$ is observed at $12.0\sigma$, which is increased to $13.7\sigma$ when any correlations due to $c$ and $\kappa_\mathrm{gal}$ have been taken into consideration. From Table~\ref{tab:pearson3} we observe that the significance of the correlation between $\Delta\mu$ and $\kappa_\mathrm{gal}$ is reduced from $1.7\sigma$ to $1.4\sigma$ when other correlations are considered. The correlation between $\kappa_\mathrm{gal}$ and $x_1$ is also much reduced. 

\citet{2010MNRAS.405.1025M} correlated the brightness of high redshift quasars with foreground galaxies to show the presence of intergalactic dust distributed up to several Mpc from galaxies. At large scales around galaxies, they infer a value of $R_V\sim4$ (which corresponds to a value of $\beta\sim5$), indicating that the color correction may be biased. However, our results show evidence for a correlation between $\kappa_\mathrm{gal}$ and $c$ at a significance of only $1.4\sigma$. 

\begin{deluxetable*}{cccc}
  \tablewidth{0pt}
  \tablecaption{Partial Correlation Coefficient, $r$ between various SNe observables considered when the effect of all other variables is removed. The significance of each correlation is also given in brackets. \label{tab:pearson3}}
    \tablehead{
    \colhead{} &
   \colhead{c} &
   \colhead{$x_1$} &
    \colhead{$\kappa_\mathrm{gal}$}
  }
  \startdata
  $\Delta \mu$ &  $-0.44 \pm 0.03 (13.7)$ &  $0.56 \pm 0.03 (20.4)$ &  $-0.06 \pm 0.04 (1.4)$ \\
c &  - &  $0.33 \pm 0.04 (9.2)$ &  $-0.06 \pm 0.04 (1.4)$ \\
$x_1$ &  - & - &  $-0.01 \pm 0.04 (0.2)$ \\
\end{deluxetable*}

\subsection{Considering lensing when estimating SN Ia distances}
\label{subsec:fixed_cosmo}

We now attempt to determine if we can improve upon the estimation of SN Ia distances using a gravitational lensing correction. To do this, we note that SN Ia magnitudes are affected by the convergence, which is correlated with the measured value of $\kappa_\mathrm{gal}$. We therefore include an additional global SN Ia parameter, $\gamma_\kappa$ in Equation~\ref{eq:basic_salt} such that

\begin{equation}
\mu = m_B - M + \alpha \,x_1 - \beta\,c + \mu_{\mathrm{corr}}(z) + \gamma_\kappa\,\kappa_{\rm gal}, 
\label{eq:modified_salt}
\end{equation}

\noindent and attempt to determine the value of $\gamma_\kappa$ simultaneously with that of $\alpha$ and $\beta$ in the cosmological fit. To be fully consistent, we include the uncertainty on our derived measurement of $\kappa_{\rm gal}$ in $\sigma_\mu$, such that $\sigma_\mu^2 = \sigma_z^2 + \sigma_{\rm fit}^2 + \sigma_{\rm int}^2 + \sigma_\kappa^2$, where $\sigma_z$ is the uncertainty on the measured redshift of each SN Ia, $\sigma_\mathrm{int}$ is the intrinsic dispersion of SNe Ia (considered in this analysis to be $\sigma_\mathrm{int}=0.12$, as described in \S\ref{sec:snedistances}) and $\sigma_\mathrm{fit}$ is the uncertainty due to the light-curve fit, which includes the uncertainties on $x_0, x_1$ and $c$. To include a low redshift anchor to the Hubble diagram, we remove our minimum redshift criteria of $z>0.2$ imposed in \S\ref{sec:estimatingkappa}, so that we have a sample of $749$ SNe Ia, each with a measured value of $\kappa_\mathrm{gal}$. To avoid any uncertainty with the absolute magnitude of an SN Ia, we additionally include a constraint for the value of $H_0=73.8\pm2.4\,\textrm{km}\,\textrm{s}^{-1}\,\textrm{Mpc}^{-1}$ from the SH0ES survey \citep{SHOES}.

To determine the value of $\gamma_\kappa$, we fix the the cosmological parameters to those given in Table~\ref{tbl:cosmo_priors} and use the Markov-Chain-Monte-Carlo (MCMC) sampler, {\it cosmoMC}, to determine the values of $\alpha, \beta$ and $\gamma_\kappa$ simultaneously.
\hspace{-4cm}
\begin{deluxetable*}{ccc}
  \tablewidth{0pt}
  \tablecaption{Priors imposed on the fitted cosmological parameters for the results in Table~\ref{tbl:cosmo_results}.\label{tbl:cosmo_priors}}
  \tablehead{
   \colhead{Parameter} &
   \colhead{Fixed Cosmology} & 
   \colhead{Fitted Cosmology}
  }
  \startdata
  $\Omega_b$ & 0.045 & 0.01,0.2 \\
  $\Omega_{DM}$ & 0.25 & -0.2,1.2 \\
  $\Omega_k$ & 0.0 & -1.0,1.0 \\
  w & -1.0 & -3.0,1.0 \\
  $\alpha$ &  0.01,0.5 & 0.01,0.5 \\
  $\beta$ & 1.0,5.0 & 1.0,5.0 \\
  $\gamma_\kappa$ & -15.0,15.0 & -15.0,15.0\\
\end{deluxetable*}

\begin{figure*}[t]
\epsscale{1.0}
\plotone{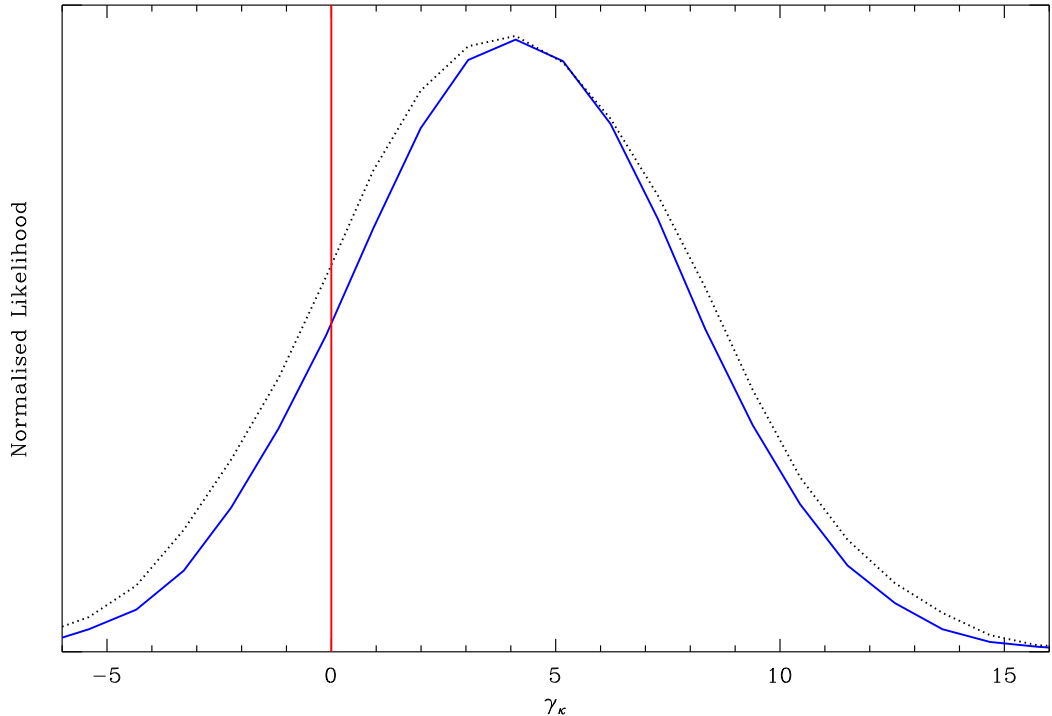}
\caption{Constraints on $\gamma_\kappa$ when a cosmology with $\Omega_m=0.3, \Omega_\Lambda=0.7$ is considered and only the SN Ia data (with $\sigma_\mathrm{int}=0.12$ mag) are used in the fit. The marginalized likelihood is shown as a solid blue line, the mean likelihood as a dotted line, while $\kappa=0$  is shown in red.}
\label{fig:fixed_cosmo}
\end{figure*}

Figure~\ref{fig:fixed_cosmo} shows the one-dimensional likelihood surface for $\gamma_\kappa$ for the sample of 749 SNe Ia used in this analysis. The marginalised likelihood is shown as a solid line, the mean likelihood as a dotted line, while $\kappa=0$  is shown in red. Marginalised parameter estimates (and $1\sigma$ uncertainties) for the three parameters, $\alpha, \beta$ and $\gamma_\kappa$, are given in Table~\ref{tbl:cosmo_results}. The recovered values of $\alpha$ and $\beta$ are consistent with the fiducial values used previously in this analysis, while a value of $\gamma_\kappa = 4.0\pm3.6$   is obtained.  No significant correlations between the 3 parameters are observed. A minimal improvement in the best-fitting $\chi^2$ is observed, with $\Delta\chi^2=1.5$. When we vary $\sigma_\mathrm{int}$ we find that the value of $\gamma_\kappa$ changes negligibly and the value of $\sigma_\mathrm{int}$ that gives a reduced $\chi^2 = 1$ does not depend on the inclusion of the $\gamma_\kappa$ parameter. 

\hspace{-4cm}
\begin{deluxetable*}{ccccccccc}
  \tablewidth{0pt}
  \tablecaption{Summary of the supernova and cosmological parameter constraints as described in \S\ref{subsec:fixed_cosmo} and \S\ref{subsec:cosmo_analysis}\label{tbl:cosmo_results}}
  \tablehead{
   \colhead{Type of Fit} &
   \colhead{Datasets} & 
   \colhead{$\Omega_m$} & 
   \colhead{w} & 
    \colhead{$\alpha$} &
    \colhead{$\beta$} &
    \colhead{$\gamma$} &
    \colhead{minimum $\chi^2$} &
    \colhead{$n_\mathrm{dof}$}
  }
  \startdata
  Fixed Cosmology & SNe Only & 0.3 & $-1.0$ & 	$0.20\pm0.01$ & $3.05\pm0.11$ & 0 & 866.05 & 747\\
  Fixed Cosmology & SNe Only & 0.3 & $-1.0$ & $0.20\pm0.01$ & $3.04\pm0.11$ & $4.0\pm3.6$ & 864.74 & 746\\
  Fitted Cosmology & SNe+CMB+BAO+$H_0$ & $0.28\pm0.02$	& $-0.99\pm0.09$	& $0.20\pm0.01$ & $3.06\pm0.11$ & 0 & 8341.86 & 8742 \\
  Fitted Cosmology & SNe+CMB+BAO+$H_0$ & $0.28\pm0.02$ 	& $-0.98\pm0.09$ & $0.20\pm0.01$	& $3.05\pm0.11$ & $4.0\pm3.6$ & 8340.29 & 8741\\
\end{deluxetable*}

\subsection{Cosmological implications} 
\label{subsec:cosmo_analysis}

Having shown that we can attempt to use the estimated value of $\kappa_\mathrm{gal}$ in our estimation of $\mu$ for SNe Ia, we now consider the implications that this additional correlation can have on the inferred cosmological parameters. In this analysis we combine the 749 SNe Ia in our sample with cosmological information from the full seven-year Wilkinson Microwave Anisotropy Probe (WMAP7) CMB power spectrum \citep{2011ApJS..192...16L}, Baryon Acoustic Oscillation (BAO) results at $z=0.2$ and $z=0.35$ determined using the SDSS DR7 main and LRG galaxy samples combined with 2dFGRS data  \citep{sdssbao}  and the SH0ES measurement of $H_0$ \citep{SHOES}. We adopt prior ranges on the cosmological parameters, as given in Table~\ref{tbl:cosmo_priors}, and using {\it cosmoMC}, constrain them simultaneously with the supernova nuisance parameters, $\alpha$ and $\beta$ and consider the implications when $\gamma_\kappa$ is included in the fit. The resulting marginalised constraints are given in Table~\ref{tbl:cosmo_results}. The central values and uncertainties on $\Omega_m$ and $w$ are unaffected by the addition of an additional parameter. The value of $\gamma_\kappa = 4.0\pm3.6$ is consistent with that found when the cosmology is held fixed. The resulting best-fit $\chi^2$ reduced by $\sim1.5$ when $\gamma_\kappa$ is included.   

\subsection{Constraining the bias of our galaxy sample} 
\label{sec:snedistances}

We now use the results from \S\ref{subsec:fixed_cosmo} to study the bias of our foreground galaxy sample. \citet{2010MNRAS.405..535J} show that to first order $\delta D_L / D_L = -\kappa$ and $\delta D_L/D_L = \Delta\mu \ln(10)/5$ for the change in $D_L$ due to lensing. Therefore, for each SN, the estimated distance modulus should have increased scatter from lensing, such that $\Delta \mu = -5\kappa/\ln{(10)}$. 

However, as described in \S\ref{sec:estimatingkappa}, we are using a sample of foreground galaxies to trace the underlying dark matter distribution along the line-of-sight to the SNe; therefore our convergence estimate will miss fluctuations in density on small scales, and will also be affected by the bias of our galaxy sample. We can write the true convergence as $\kappa_{\rm true} = \kappa_{\rm win} + \kappa_{\rm ex}$, where $\kappa_{\rm win}$ is the convergence averaged in our aperture and $\kappa_{\rm ex}$ is the difference between this and the true convergence; $\kappa_{\rm win}$ is only very weakly correlated with $\kappa_{\rm ex}$. Assuming a linear bias $b$, the windowed convergence is related to our estimator by $\kappa_{\rm win} = \kappa_\mathrm{gal}/b$. Then the distance moduli of the SNe Ia will be altered so that

\begin{equation}
\mathrm{\mu}_\mathrm{obs} = \mathrm{\mu}_{\mathrm{true}} + 5\,\kappa_\mathrm{gal}/b\,\ln{(10)} + 5 \kappa_{\rm ex} / \ln{(10)}
\label{eq:lensingeffect}
\end{equation}

where $\mu_{\rm true}$ is the unlensed distance modulus. The third term will add extra scatter, but is not probed by our method; the second term corresponds to the final term of Equation {\ref{eq:modified_salt}}. Therefore combining Equations~\ref{eq:modified_salt} and~\ref{eq:lensingeffect}, we anticipate a value of $\gamma_\kappa = 5/b\,\ln{10} \simeq 2.17/\,b$. 

Our sample of SDSS foreground galaxies is comprised of three major sub-samples with different selection criteria (\S\ref{subsec:galdata}). As such, it is difficult to accurately estimate the bias ($b$) for such a merged sample with respect to the underlying matter distribution. However, \citet{2005PhRvD..71d3511S} estimate a value of $b = 0.99\pm0.07$ for the SDSS MGS, which when combined with the {\it Southern} program, which has a similar selection criteria to the MGS, comprise $72\%$ of the galaxies in our sample, while \citet{2007MNRAS.378.1196K} estimate a value of $b=1.87\pm0.07$ for LRGs from SDSS DR3. Combining these estimates produces an anticipated value of $b=1.24\pm0.10$ for our foreground galaxy sample. Our measured value of $\gamma_\kappa = 4.0\pm3.6$, gives $b = 0.54\pm0.48$, in excellent agreement with the value measured for the SDSS MGS sample and consistent with the value for of combined sample at $1.4\sigma$.

\section{Conclusions} 
\label{sec:conclusions} 

In this paper, we have introduced a method to measure the effect of weak gravitational lensing on SN Ia distances and include this information when determining the cosmological parameters. To demonstrate this scheme, we use a sample of 608 SNe Ia with spectroscopic host galaxy redshifts from the SDSS-II SNe and BOSS surveys \citep{2013ApJ...763...88C}, to $z<0.6$.  We find,
\begin{itemize}
\item At a significance of $1.7\sigma$ there exists a correlation between Hubble diagram residuals and the measured lensing convergence, $\kappa_\mathrm{gal}$ along lines-of-sight to the SN Ia positions. This correlation is consistent with the expected lensing signal such that SNe Ia along lines of sight with $\kappa_\mathrm{gal}>0$ are brighter, after correction, than those with  $\kappa_\mathrm{gal}<0$. This result is observed when various aperture radii are considered, but peaks at an averaging radius for $\kappa_\mathrm{gal}$ of $12^\prime$ (Appendix~\ref{app:angular}). The significance of our correlation is comparable that found by \citet{2010A&A...514A..44K} and \citet{2010MNRAS.405..535J} using the SNLS dataset and \citet{Jonsson:2006eu} using 26 SNe Ia from the GOODS field.
\item For this dataset, we find a strong correlation (at over $8\sigma$) between $\Delta \mu$ and other SNe Ia observables, $x_1$ and $c$ {\it after} a linear correction for these variables has been applied. 
\item We have studied whether the correlation between $\Delta \mu$ and $\kappa_\mathrm{gal}$ can be explained through correlations between other SN Ia observables, and find that when correlations including those between $x_1$ and $c$ are considered, the inferred lensing signal is observed at $1.4\sigma$.
\item We show that $\kappa_\mathrm{gal}$ and $c$ are correlated at a significance of only $1.4\sigma$, indicating that the correlation between $\kappa_\mathrm{gal}$ and Hubble diagram residuals is not caused by dust. 
\item To improve the standardisation of SNe Ia, we consider an additional parameter, $\gamma_\kappa$, when determining the distance to a SNe Ia, using the SALT2 formalism, such that the distance is linearly related to $\kappa_\mathrm{gal}$. We constrain this parameter simultaneously with other SN Ia global parameters, $\alpha$ and $\beta$, and find a value of $\gamma_\kappa=4.0\pm3.6$, fully consistent with the expected result from lensing.  When we vary $\sigma_\mathrm{int}$ we find that the value of $\gamma_\kappa$ changes negligibly and the value of $\sigma_\mathrm{int}$ that gives a reduced $\chi^2 = 1$ does not depend on $\gamma_\kappa$ indicating that the observed dispersion in $\mu$ is not primarily caused by lensing. 
\item We combine our SN Ia dataset with data from WMAP7, SDSS BAO measurements and $H_0$ measurements from HST to constrain the standard cosmological model, when SN Ia lensing is included in the cosmological analysis. We find that the inclusion of an additional parameter based on $\kappa_\mathrm{gal}$ does not affect the central values or uncertanties on $\Omega_m$ and $w$.
\item We compare our value of $\gamma_\kappa=4.0\pm3.6$ to that anticipated assuming a linear bias, and find $b = 0.54\pm0.48$ for our sample of foreground galaxies, entirely consistent with that found by galaxy clustering analyses for the MGS.  
\end{itemize} 

Our obtained correlations and constraints on $\gamma_\kappa$ are not statistically significant due to the limited redshift range covered by the BOSS sample and the relatively small number of SNe Ia in our dataset. However, forthcoming SNe surveys, such as the Dark Energy Survey \citep{dessn}, will obtain well-measured light-curves for thousands of SNe to $z>1$, and thus be dominated by systematic uncertainties. An understanding of the lensing signal expected by these surveys is important to produce the most accurate constraints on the equation of state of Dark Energy, $w$.  

\acknowledgments

\section*{Acknowledgements}

Please contact the authors to request access to research materials discussed in this paper. MS and RM are supported by the South African Square Kilometre Array Project and the South African National Research Foundation. DB, RN and RM are supported by the UK Science \& Technology Facilities Council (grant nos. ST/H002774/1 and ST/K0090X/1). The work of CC and BB was supported by the South African National Research Foundation.  This work was partially support by STFC grant ST/K00090X/1 and a Royal Society-NRF International Exchange Grant. CS is funded by a NASA Postdoctoral Program fellowship through the Jet Propulsion Laboratory, California Institute of Technology. Computations were done on the Sciama High Performance Compute (HPC) cluster which is supported by the ICG, SEPNet and the University of Portsmouth. MS thanks Russell Johnston for insightful comments. 

Funding for the SDSS and SDSS-II has been provided by the Alfred P. Sloan Foundation, the Participating Institutions, the National Science Foundation, the U.S. Department of Energy, the National Aeronautics and Space Administration, the Japanese Monbukagakusho, the Max Planck Society, and the Higher Education Funding Council for England. The SDSS Web Site is http://www.sdss.org/.

The SDSS is managed by the Astrophysical Research Consortium for the Participating Institutions. The Participating Institutions are the American Museum of Natural History, Astrophysical Institute Potsdam, University of Basel, University of Cambridge, Case Western Reserve University, University of Chicago, Drexel University, Fermilab, the Institute for Advanced Study, the Japan Participation Group, Johns Hopkins University, the Joint Institute for Nuclear Astrophysics, the Kavli Institute for Particle Astrophysics and Cosmology, the Korean Scientist Group, the Chinese Academy of Sciences (LAMOST), Los Alamos National Laboratory, the Max-Planck-Institute for Astronomy (MPIA), the Max-Planck-Institute for Astrophysics (MPA), New Mexico State University, Ohio State University, University of Pittsburgh, University of Portsmouth, Princeton University, the United States Naval Observatory, and the University of Washington.

Funding for SDSS-III has been provided by the Alfred P. Sloan Foundation, the Participating Institutions, the National Science Foundation, and the U.S. Department of Energy Office of Science. The SDSS-III web site is http://www.sdss3.org/.

SDSS-III is managed by the Astrophysical Research Consortium for the Participating Institutions of the SDSS-III Collaboration including the University of Arizona, the Brazilian Participation Group, Brookhaven National Laboratory, University of Cambridge, Carnegie Mellon University, University of Florida, the French Participation Group, the German Participation Group, Harvard University, the Instituto de Astrofisica de Canarias, the Michigan State/Notre Dame/JINA Participation Group, Johns Hopkins University, Lawrence Berkeley National Laboratory, Max Planck Institute for Astrophysics, Max Planck Institute for Extraterrestrial Physics, New Mexico State University, New York University, Ohio State University, Pennsylvania State University, University of Portsmouth, Princeton University, the Spanish Participation Group, University of Tokyo, University of Utah, Vanderbilt University, University of Virginia, University of Washington, and Yale University.

\bibliographystyle{natbib}
\bibliography{mybibliography}	

\appendix

\section{Choice of Aperture}
\label{app:angular}

For an individual line-of-sight, we require a method for determining an aperture within which to count galaxies, in order to describe the overdensity along that line-of-sight. In this appendix we investigate  two approaches involving fixed apertures around each SNe; we do not consider adaptive apertures here.    

First, we consider an aperture of fixed angular radius (e.g. \citep{2004MNRAS.351.1387W}). This is easy to define, but suffers from the fact that the transverse physical separation between a SN Ia and a foreground galaxy is a function of redshift. Secondly, we consider an aperture of fixed physical scale so that a galaxy of a fixed physical transverse  separation from the SN light--of--light is included. In this second case, we need to define the cosmological background (in particular the Hubble constant, $H_0$). For this test, we assume a flat Universe with $\Omega_m=0.3$ and $H_0 = 73.8 \textrm{km}\,\textrm{s}^{-1}\,\textrm{Mpc}^{-1}$ \citep{SHOES}.

We test the robustness of these two measurements against each other in Figure~\ref{fig:kappavkappa} (top panel). In this case, we have compared a fixed angular aperture of 12 arcminutes and a fixed physical aperture of $3\textrm{Mpc}$. The two estimates are strongly correlated, with $\rho=0.86$, indicating that either of these aperture measures will act as a similar proxy for overdensity. We considered apertures from 8-15 arcminutes and 2-10$\textrm{Mpc}$ and found a similar level of consistency. 

\section{Optimal Aperture Size}
\label{app:opt_ap}

\begin{figure*}[t]
\epsscale{1.0}
\plotone{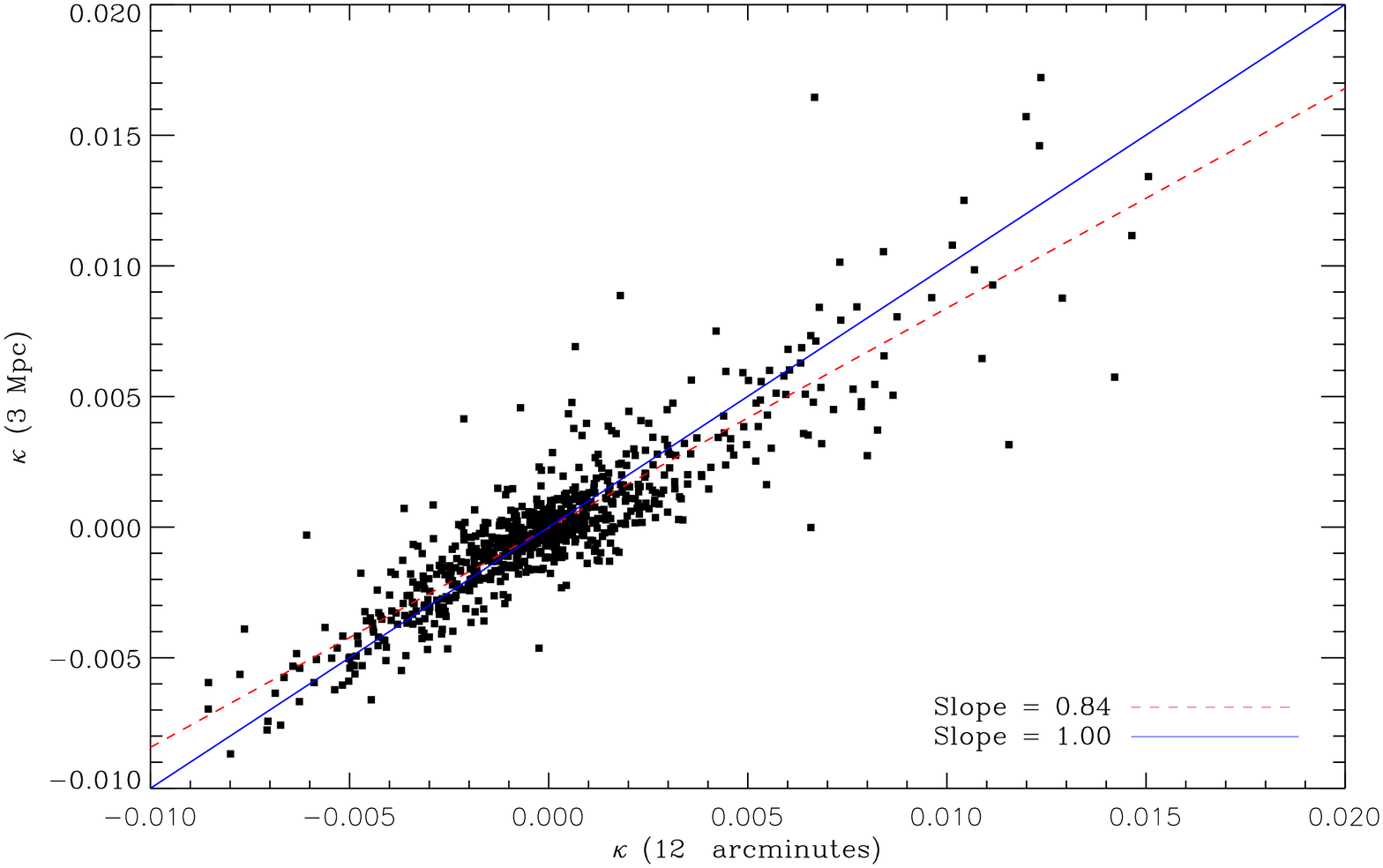}
\plotone{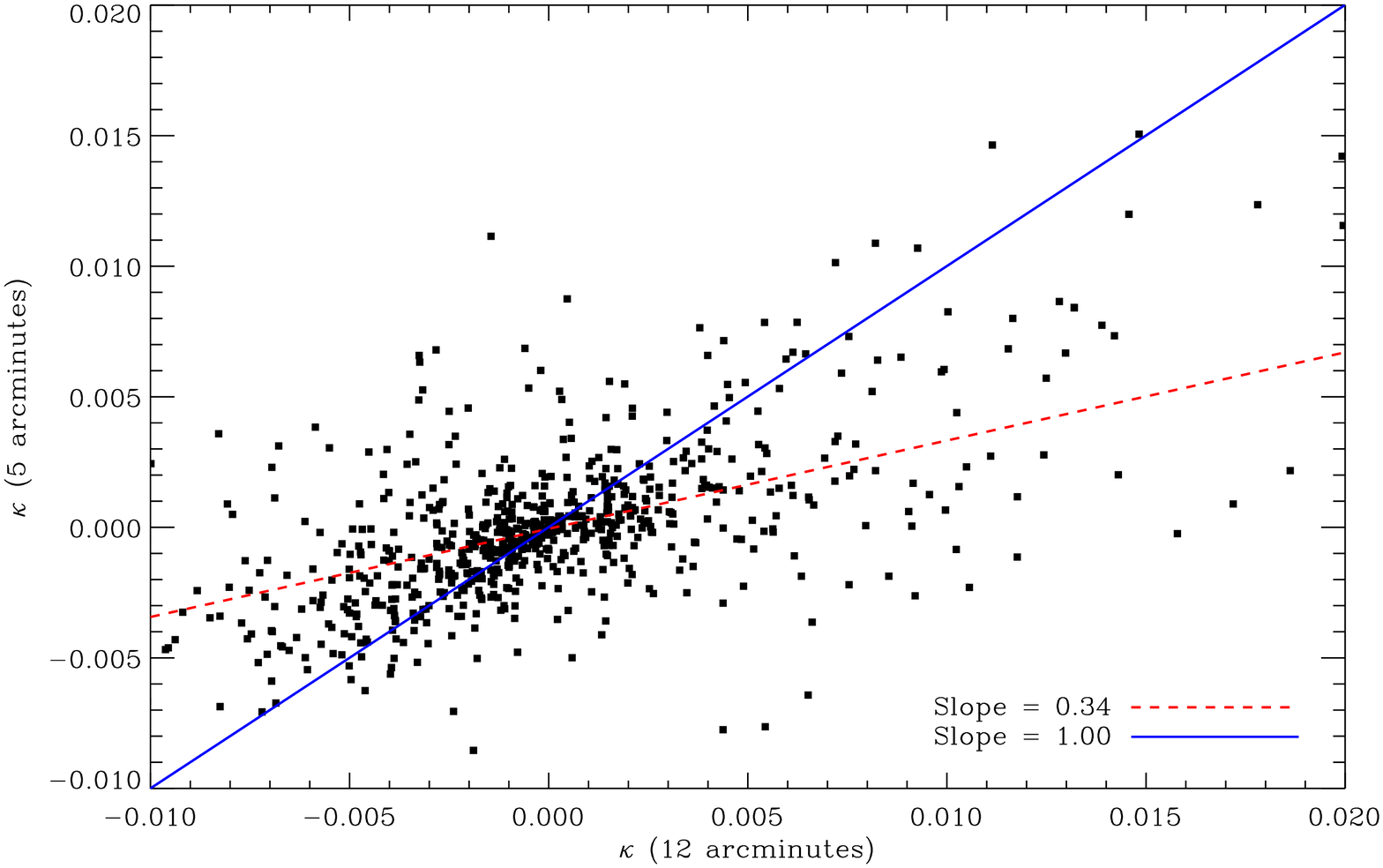}	
\caption{\emph{Top: } $\kappa_\mathrm{gal}$ determined using a fixed angular aperture of $12^\prime$ compared to the case when a fixed physical aperture of $3\textrm{Mpc}$ is considered. \emph{Bottom:} $\kappa_\mathrm{gal}$ determined for an aperture of $5^\prime$ compared to that of $12^\prime$. In both cases, a line of best-fit is shown in red, with a one-to-one correlation plotted in blue. \label{fig:kappavkappa}}
\end{figure*}

Having shown that our estimate of $\kappa_\mathrm{gal}$ is insensitive to the method used to determine the aperture, we now consider how the aperture size considered affects the distribution of $\kappa_\mathrm{gal}$. Smaller apertures are likely to be dominated by individual galaxies while larger radii will trace the mean matter distribution. The bottom panel of Figure~\ref{fig:kappavkappa} shows the estimated value of $\kappa_\mathrm{gal}$ for two aperture radii; $5^\prime$ and $12^\prime$, respectively. The two distributions are strongly correlated, with $r=0.57$, indicating that at these scales the recovered value of $\kappa_\mathrm{gal}$ is robust to the aperture size considered. 

We observe that in each case, $\kappa_\mathrm{gal} \sim \mathcal{O}(0.01)$, predicting a $1\%$ lensing signal. 

\begin{figure*}[t]
\epsscale{1.0}
\plotone{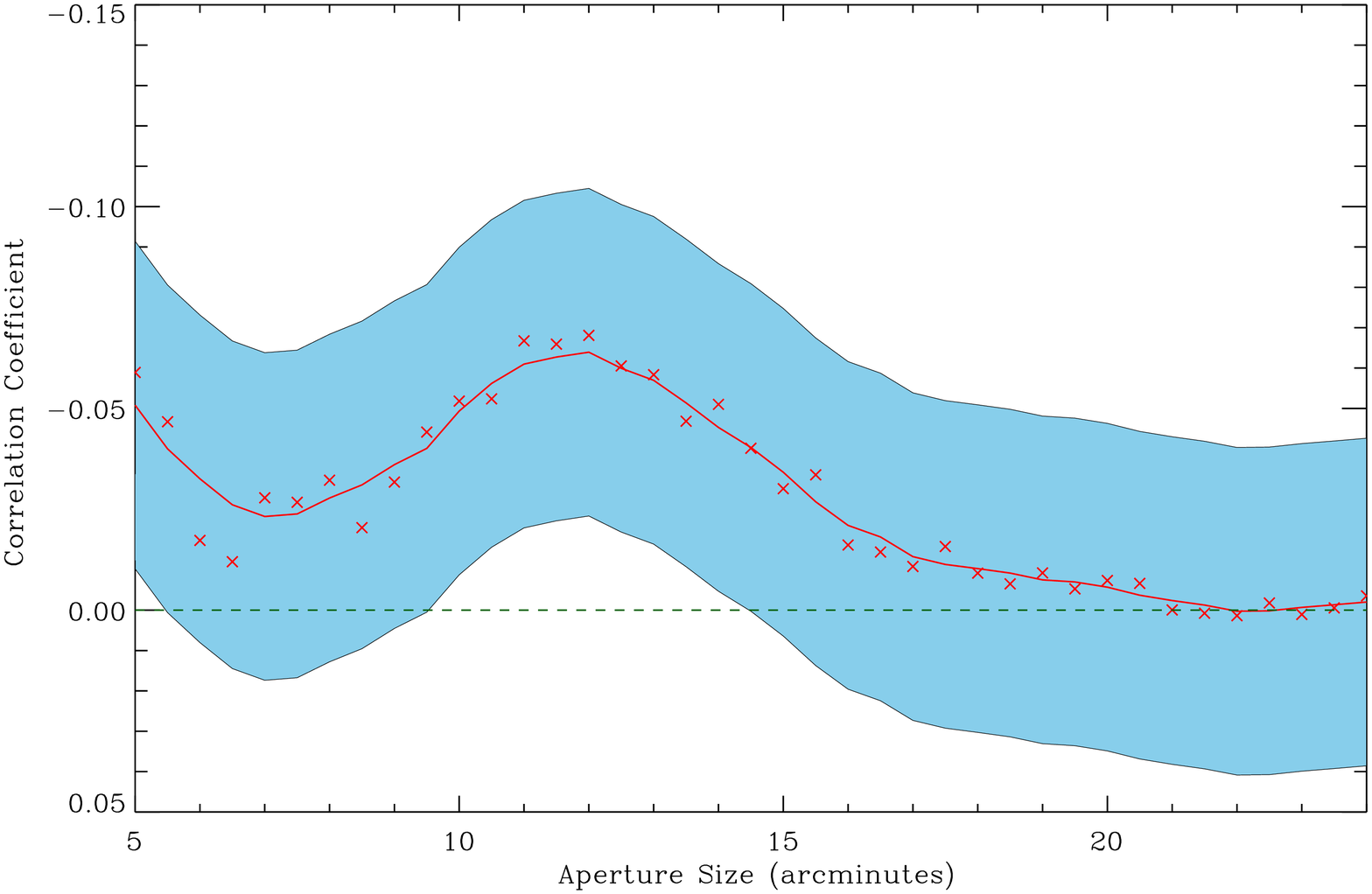}		
\caption{The Spearman's rank correlation coefficient, $\rho$ as a function of aperture radii considered, when a fixed angular size aperture is considered. The data has been smoothed, and is overplotted in red. The uncertainty in the correlation coefficient is shown as a blue band. 
\label{fig:corr_v_separation}}
\end{figure*}

Next, we consider the optimal aperture size for our data and measurement. Using the Spearman correlation coefficient and the data shown on Figure~\ref{fig:kappa_v_resid}, we re-calculate $\rho$ as a function of the aperture size used to calculate $\kappa_\mathrm{gal}$, which is then correlated with the Hubble residuals ($\mu_{obs} - \mu_{cosmo}$). We fix the cosmological and supernova nuisance parameters as described in \S\ref{sec:snedistances}. In Figure~\ref{fig:corr_v_separation}, we show the value of $\rho$ as a function of aperture size, and witness a clear ``bump" in the strength of the correlation between aperture sizes of $10$ to $15$ arcminutes, with the maximum near 12 arcminutes. This observed signal is consistent with the expected lensing prediction, with a negative correlation indicating that SNe with $\kappa_\mathrm{gal}>0$ are marginally brighter than those with  $\kappa_\mathrm{gal}<0$, after correction. The reduction in the signal at larger aperture radii is due to these apertures picking up other structures which are not causing the SNe lensing.

\section{Effect of a minimum redshift limit}
\label{app:zmin}

\begin{figure*}[t]
\epsscale{1.0}
\plotone{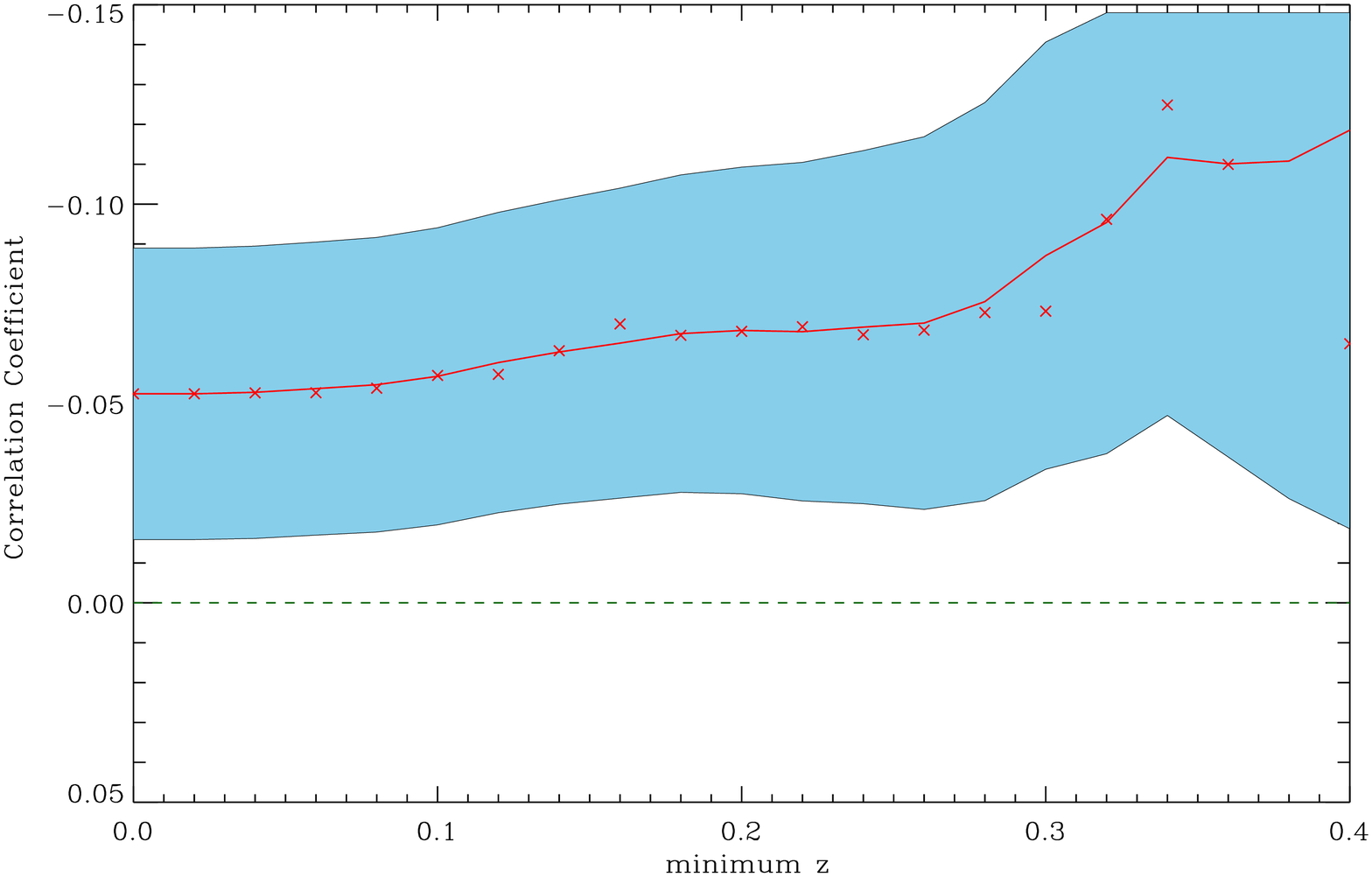}		
\caption{The Spearman's rank correlation coefficient, $\rho$ as a function of minimum redshift considered, when a fixed angular size aperture of 12 arcseconds is considered.  The uncertainty in the correlation coefficient is shown as a blue band.
\label{fig:corr_v_redshift}}
\end{figure*}

In \S\ref{sec:estimatingkappa} we enforced that only SNe Ia with $z>0.2$ are included in our fiducial sample. In this appendix we consider the implications that this cut has on our results. Using the Spearman correlation coefficient discussed and the data shown on Figure~\ref{fig:kappa_v_resid}, we re-calculate $\rho$ as a function of the minimum redshift used in the sample, which is then correlated with the Hubble residuals ($\mu_{obs} - \mu_{cosmo}$). We fix the cosmological and supernova nuisance parameters as described in \S\ref{sec:snedistances}. In Figure~\ref{fig:corr_v_redshift}, we shows the value of $\rho$ as a function of minimum redshift, and observe a clear increase in the signal with increasing redshift, indicating that SNe Ia at higher redshifts are more sensitive to the lensing signal, as expected. However, with the increasing minimum redshift, the size of our the resulting sample decreases, increasing the inferred uncertainties. A signal is observed independent of the redshift cut considered. 

\end{document}